\documentclass[aps, prresearch, superscriptaddress, notitlepage, reprint, longbibliography]{revtex4-1}
\usepackage[english]{babel}
\usepackage{amsmath,amsthm}
\usepackage{amsfonts}
\usepackage[pdfborder={0 0 0}, colorlinks=true, urlcolor=blue, linkcolor=blue, citecolor=blue]{hyperref}
\usepackage{color}
\usepackage{graphicx}
\usepackage{float}
\usepackage{subfigure}
\usepackage{lipsum}

\theoremstyle{definition}

\theoremstyle{remark}

\begin{document}

\title{Nonperturbative leakage elimination for a logical qubit encoded in a mechanical oscillator}
\author{Shasha Zheng}
\affiliation{State Key Laboratory of Mesoscopic Physics, School of Physics, Frontiers Science Center for Nano-optoelectronics, $\&$ Collaborative Innovation Center of Quantum Matter, Peking University, Beijing 100871, China}
\affiliation{Department of Theoretical Physics and History of Science, The Basque Country University (EHU/UPV), 48008, Spain}

\author{Qiongyi He}
\email{qiongyihe@pku.edu.cn}
\affiliation{State Key Laboratory of Mesoscopic Physics, School of Physics, Frontiers Science Center for Nano-optoelectronics, $\&$ Collaborative Innovation Center of Quantum Matter, Peking University, Beijing 100871, China}
\affiliation{Beijing Academy of Quantum Information Sciences, Beijing 100193, China}
\affiliation{Collaborative Innovation Center of Extreme Optics, Shanxi University, Taiyuan 030006, China}

\author{Mark S. Byrd}
\email{mbyrd@siu.edu}
\affiliation{Department of Physics, Southern Illinois University, Carbondale, Illinois 62901-4401, USA}

\author{Lian-Ao Wu}
\email{lianaowu@gmail.com}
\affiliation{Department of Theoretical Physics and History of Science, The Basque Country University (EHU/UPV), 48008, Spain}
\affiliation{IKERBASQUE, Basque Foundation for Science, 48011 Bilbao, Spain}

\begin{abstract}
Continuous-variable (CV) systems are attracting increasing attention in the realization of universal quantum computation. Several recent experiments have shown the feasibility of using CV systems to, e.g., encode a qubit into a trapped-ion mechanical oscillator and perform logic gates [Nature 566, 513-517 (2019)]. The essential next step is to protect the encoded qubit from quantum decoherence, e.g., the motional decoherence due to the interaction between a mechanical oscillator and its environment.  Here we propose a scheme to suppress quantum decoherence of a single-mode harmonic oscillator used to encode qubits by introducing a nonperturbative leakage elimination operator (LEO) specifically designed for this purpose. Remarkably, our nonperturbative LEO can be used to analytically derive exact equations of motion without approximations. It also allows us to prove that the effectiveness of these LEOs only depends on the integral of the pulse sequence in the time domain, while details of the pulse shape does not make a significant difference when the time period is chosen appropriately. This control method can be applied to a system at an arbitrary temperature and arbitrary system-bath coupling strength which makes it extremely useful for general open quantum systems.
\end{abstract}
\maketitle

\section{INTRODUCTION}
Decoherence, which is inevitable due to the coupling with the surrounding environment, is regarded as the main barrier to modern quantum technologies such as quantum computing. Suppression of decoherence is a long-lasting challenge, and has attracted considerable attention both for investigations of fundamental theories in quantum mechanics and for practical applications in nanoengineering~\cite{Maximilian2007, Suter2016Review}. Fortunately, an open-loop quantum control method named dynamical decoupling, originating from spin-echo effect in nuclear magnetic resonance experiments~\cite{Hahn1950, Carr1954},
provides a universal and effective method in suppressing decoherence.  The effectiveness of these controls has been confirmed both in theory~\cite{Viola1998, Viola1999, Vitali1999, Zanardi2000, YaoQi2019} and abundant experiments largely focused on two-level systems such as electron or nuclear spins in solid-state system~\cite{Morton2008, DuJiangfeng2009}, semiconductor quantum dots~\cite{Barthel2010, Bluhm2011}, diamond nitrogen-vacancy centers~\cite{Ryan2010, deLange2010, XuXiangkun2012, Bar2013}, atomic ensembles~\cite{Biercuk2009, BiercukPRA2009, Sagi2010}, superconducting qubits~\cite{Bylander2011, Pokharel2018}, and photonic qubits in a ring cavity~\cite{Damodarakurup2009}. However, the quantum control theory based on leakage elimination operation has remained out of reach for continuous-variable (CV) systems, such as a harmonic oscillator.

Latest experiments have demonstrated the feasibility of using CV systems in encoding qubits into CV systems and performing logic gates~\cite{Fluhmann2019,Gao2019,Lescanne2020}. The higher-dimensional space benefits from large information storage and hardware-efficient quantum error correction protocols, which allows detection of small  shift errors and correction without disturbing the information stored in the state. Specifically, a trapped-ion mechanical oscillator was proposed to realize encoded qubits~\cite{Fluhmann2019} based on the Gottesman-Kitaev-Preskill (GKP) code. The GKP code is an error-correcting code useful for detecting and correcting small (classical) errors contained in phase space displacements~\cite{GKP2001,Albert2018,Terhal2015}. Thus, the natural next step is to protect the encoded qubit from quantum decoherence due to the interaction between the CV system and its quantum environment. 
 
Leakage elimination operators (LEOs)~\cite{Byrd/Lidar:02,Wu/etal:02,Byrd/etal:05} can be used to suppress errors by eliminating or reducing the interaction between the system and bath. LEOs specifically focus on eliminating leakage from the encoded subspace to other states of the system. Although the suppression of decoherence of a damped harmonic oscillator has been studied by using suitably tailored external forcing~\cite{Vitali1999} based on the so-called `bangbang' (BB) controls, it is essentially a perturbative theory by assuming that the pulses are extremely strong so that the system-reservoir interaction Hamiltonian can be neglected during pulsing~\cite{JingJunPRA2013}. This is quite unrealistic for most experiments. Therefore, it has become a worthwhile objective to develop nonperturbative versions of LEO theory for a physical model with realistic system-bath interactions where an arbitrary pulse sequences can be employed in the control process.

The general idea of LEO suppression is to modify the interaction Hamiltonian by successive pulse controls so that the average effect of the unwanted environment is eliminated~\cite{Viola1998, Viola1999, Vitali1999, Zanardi2000}. This can be illustrated as follows.  The Hamiltonian for the desired system and environment can be written as $H=H_S+H_B+H_{int}$, where $H_S,~H_B,~H_{int}$ represents, respectively, the Hamiltonian of the system, environment, and the interaction between the system of interest and the bath. If an operator $R_L$ satisfies the symmetry condition $\{R_L,H_{int}\}=0,~[R_L,H_{S}]=0$, and $[R_L,H_{B}]=0$, it is an example of an LEO. Although other more complicated pulse sequences exist, this is the most efficient and often easiest to implement experimentally.  As $\lim_{m\rightarrow \infty}(e^{-iHt/m}R_L^\dagger e^{-iHt/m}R_L)^m=e^{-iH_St}e^{-iH_Bt}$, one can entirely remove the influence of the environment by employing a well designed successive set of time-dependent pulses acting only on the system. For a single-mode harmonic oscillator, we aim to eliminate leakage to other energy levels. In this case $H_S=\omega_0a^\dagger a$, and an LEO is given by $R_L={\exp}(-i\pi a^\dagger a)$ according to the symmetry requirements, as long as the interaction Hamiltonian is of the form $H_{int}=\sum_kg_kB_k(a+a^\dagger)$ and no matter the specific form of the environment. And this LEO can be obtained by operating a controllable Hamiltonian $H_{LEO}=C(t) a^\dagger a$ on the desired system for a time $\tau$, i.e., $R_L={\rm exp}(-iH_{LEO}\tau)$. Note that $B_k=b_k+b^\dagger_k$ when the harmonic oscillator is coupled with a reservoir described by a variety of independent oscillators $b_k$ as considered in this paper. Note that the method proposed here is also feasible  to suppress the decoherence when the harmonic oscillator interacts with other forms of environment, e.g. a spin bath of $N$ independent spin-$\frac{1}{2}$ particles, in which the interaction Hamiltonian reads $H_{int}=\sum_kg_k(a+a^\dagger)\sigma_z^{(k)}$~\cite{Schlosshauer2008}, then $B_k=\sigma_z^{(k)}$ with $\sigma_z^{(k)}$ being the Pauli matrix of the $k$-th spin particle.

Here, we investigate the nonperturbative dynamical equation of a single-mode harmonic oscillator with time-dependent frequency via the Heisenberg-Langevian method. Different from the widely used quantum-state-diffusion (QSD) equation which is only exactly valid for zero temperature~\cite{JingJunPRA2013, JingJunPRL2015}, the present control method is valid for arbitrary temperature and arbitrary system-bath coupling strength.  This makes it possible to suppress the decoherence of an open system surrounded by high-temperature and high coupling strength environment. We derive the exact equations of motion without any approximations, and find that the expectation value of the annihilation operator can be perfectly preserved when implementing successive external pulses on the desired system. We further study the influence of important factors on the effectiveness of LEO control and compare the performance of three different types of pulses. 

The paper is organized as follows. In Sec.~\ref{General_formalism} we introduce the model and derive the exact equation of motion of a single-mode harmonic oscillator. In Sec.~\ref{Results} we present the results for the suppression of decoherence of a harmonic oscillator by applying three different types of pulses. We show that the effectiveness of quantum control is determined by the integrals of the pulse sequences, while is irrelevant to details of the pulse shape when the time period is relatively small. Finally, we summarize
our results in Sec.~\ref{Conclusion}.

\section{General formalism}\label{General_formalism}
Consider a system with a single-mode harmonic oscillator ($a$ with frequency $\omega_a$) interacting with a reservoir described by a variety of independent bosonic oscillators ($b_j$ with frequencies $\omega_j$).  The Hamiltonian after the rotating wave approximation is given by
\begin{eqnarray} \label{H}
H/\hbar=\omega_a(t)a^\dagger a +\sum_j\omega_j b_j^\dagger b_j +\sum_j g_j (b_j^\dagger a+b_j a^\dagger),
\end{eqnarray}
where $g_j$ is the coupling strength between the oscillator and the reservoir. We will assume that the product of the oscillator's mass and frequency is constant \cite{Mandal2017}, i.e., $m(t)\omega_a(t)=\text{constant}$, indicating the annihilation and creation operators are not explicitly time-dependent. The Heisenberg equations of motion are $\dot{a}=i[H,a]/\hbar=-i\omega_a(t)a-i \sum_j g_j b_j$, $\dot{b}_j=i[H,b_j]/\hbar=-i\omega_j b_j-i g_j a$, respectively. The equations for the reservoir operators can be directly integrated, which are $b_j(t)=b_j(0)e^{-i\omega_j t}-i g_j \int_0^t dt' a(t') e^{-i \omega_j (t-t')}$. $a(t)$ can be transformed to the slowly varying annihilation operator $A(t)=e^{i\int_0^t  \omega_a(s) ds}a(t)$ which satisfies
\begin{eqnarray}\label{eq_A}
\dot{A}&=&-\int_0^t dt' G(t-t') e^{-i\int_t^{t'} \omega_a(s)ds} A(t')+F(t).
\end{eqnarray}
Here, $G(t)=\int d\omega J(\omega) e^{-i\omega t}$ is the Fourier transformation of the interaction spectrum intensity $J(\omega)=\sum_j |g_j|^2 \delta(\omega-\omega_j)$. When $e^{-i\int_t^{t'} \omega_a(s)ds}$ is a fast oscillation term and $G(t-t')$ is a slowly varying function, the first integral term in Eq.~(\ref{eq_A}) tends to zero according to Riemann-Lebesgure lemma~\cite{JingJunPRA2013, JingJunPRA2014, JingJunPRL2015, WangPRA2018Dec}. Then Eq.~(\ref{eq_A}) reduces to $\dot{A}=F(t)$, where $F(t)=-i\sum_j g_jb_j(0)e^{\int_0^t i[\omega_a(s)-\omega_j]ds}$ denotes the noise operator of the environment and thus determines the evolution involving the harmonic oscillator operator. Note that the Heisenberg-Langevian approach works for arbitrary temperature, which is different from the QSD equation for zero temperature~\cite{JingJunPRA2013, JingJunPRL2015}. However, $\int_0^tF(t)dt=0$ if $e^{i\int_0^{t} [\omega_a(s)-\omega_j]ds}$ is quickly oscillating. Therefore, $A(t)=A(0)$, illustrating that the desired harmonic oscillator system is decoupled from the environment when the frequency of the oscillator is appropriately chosen.
 
 The system is assumed to have an initial state $|\alpha_0\rangle\langle \alpha_0|\bigotimes\prod_j\rho_j$, where the reservoir is in thermal equilibrium $\rho_j=\sum_{n_j}|n_j\rangle\langle n_j|e^{-\beta n_j\hbar\omega_j}/z$ with $z=\sum_{n_j=0}^{\infty} e^{-\beta n_j\hbar\omega_j}$, $\beta=1/(k_B T_e)$, $k_B$ denotes the Boltzmann constant, and $T_e$ is the temperature of the environment. We then investigate the evolution of the expectation value $\alpha(t)=\langle A(t)\rangle$ which is independent of reservoir temperature simply because the expectation value of $F(t)=-i\sum_j g_jb_j(0)e^{\int_0^t i[\omega_a(s)-\omega_j]ds} $ is zero as $tr(\rho_j b_j(0))=0$. Therefore, its evolution satisfies 
\begin{equation}\label{alpha_A}
\dot{\alpha}(t)=-\int_0^t dt' G(t-t') e^{-i\int_t^{t'} \omega_a(s)ds} \alpha(t').
\end{equation}
Here, we consider a non-Markovian environment corresponding to the Ornstein-Uhlenbeck process with the exponential decay correlation function $G(t)=e^{-\gamma_0 |t|}\Gamma \gamma_0/2$,
where $1/\gamma_0$ characterizes the memory time of the environment and $\Gamma$ denotes the coupling strength between the system and bath. In the limit $\gamma_0\rightarrow \infty$ the system evolves under a Markovian environment with white noise. 
\begin{figure}
\includegraphics[width=0.48\columnwidth]{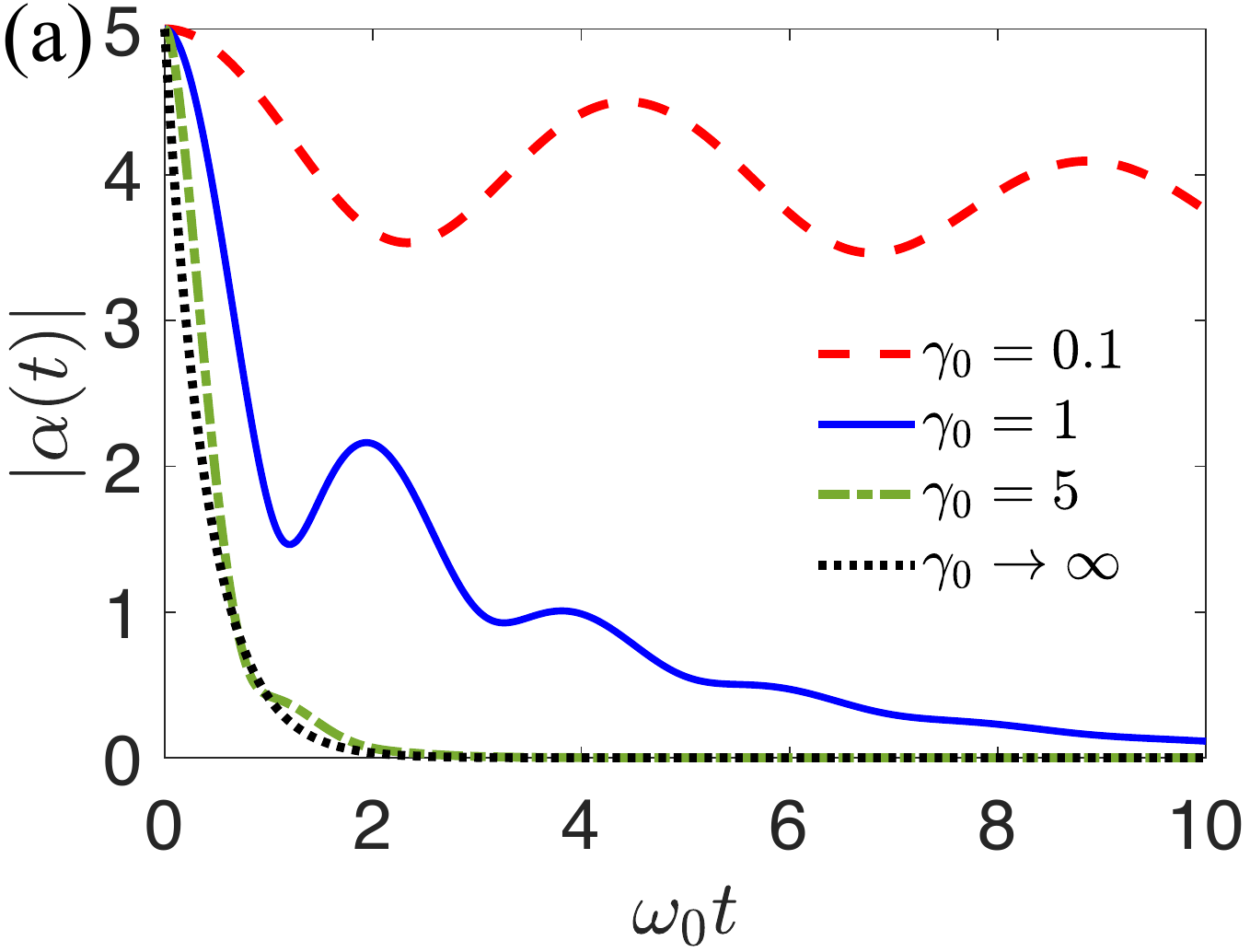}
\includegraphics[width=0.48\columnwidth]{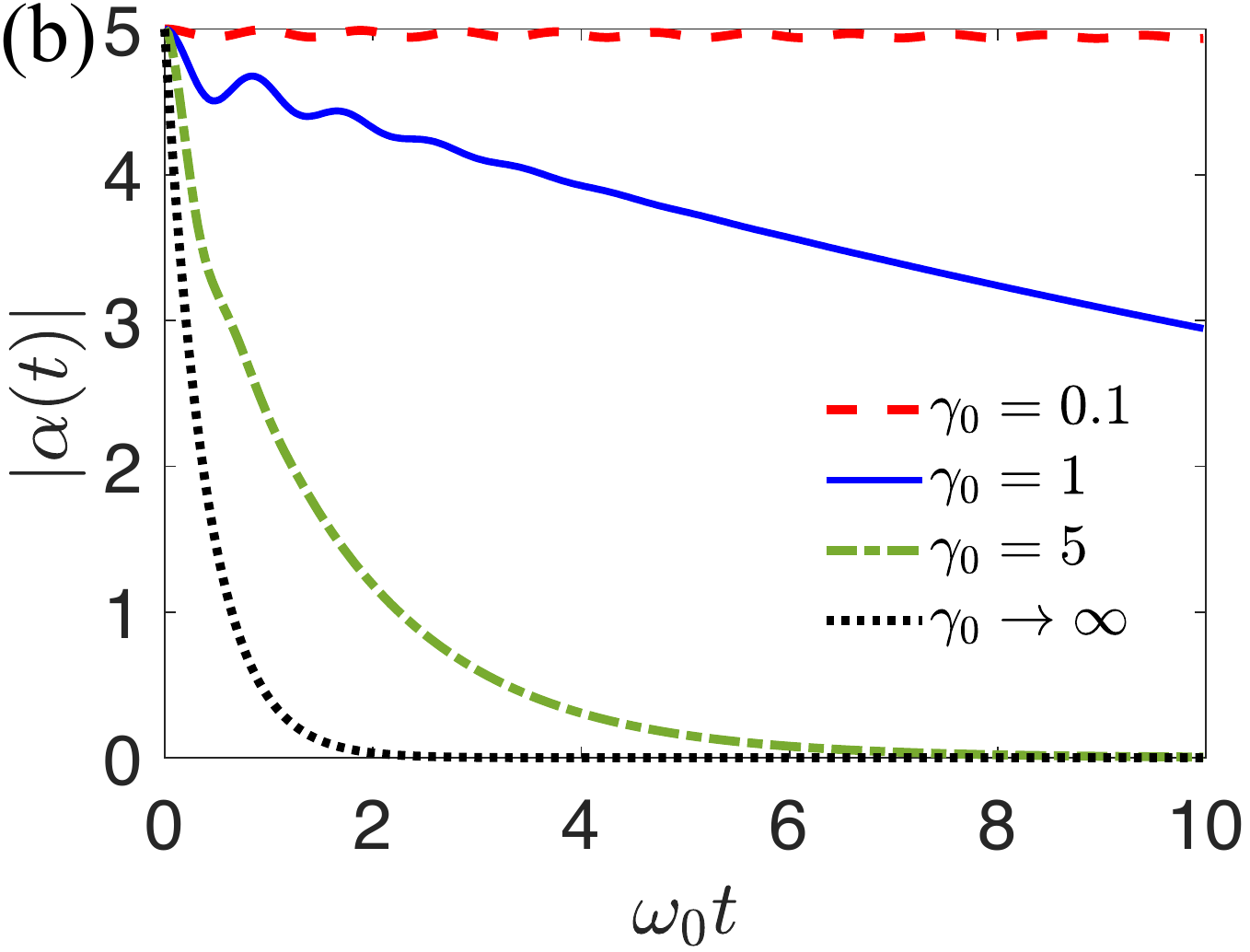}
\caption{The coherent amplitude $|\alpha(t)|$ as a function of time for different memory times $\gamma_0$ without control (a) and with regular rectangular control (b). Other parameters are $\alpha(0)=5,~\omega_0=1, ~\Gamma=5,~\omega_1=8,~T=0.05,~\Delta/T=0.7$. For simplicity, these parameters are expressed in units of the bare frequency $\omega_0$, and the parameters involving time, such as $T,~\Delta,~t$, are written in the units of $1/\omega_0$.}
\label{Fig1}
\end{figure}

\section{results}\label{Results}
\subsection{Oscillator motion without quantum control}
Without quantum control, the frequency of the oscillator is constant, i.e., $\omega_a(t)=\omega_0$, where $\omega_0$ is the bare frequency of the oscillator. The general solution of the coherent amplitude can be found by direct computation as in Appendix~\ref{no_control}. It is worth noting that when $\gamma_0\rightarrow \infty$, the correlation function reduces to $G(t)=\delta(t)\Gamma/2$. Then the coherent amplitude of the oscillator of interest is $\alpha(t)=\alpha_0 e^{-\Gamma t/2}$, indicating that the ideal Markovian process is uncontrollable. 

Figure~\ref{Fig1}(a) shows the time dependence of the coherent amplitude $|\alpha(t)|$ for different $\gamma_0$, which can be used to find the boundary between Markovian and non-Markovian dynamics through increasing $\gamma_0$. It can be seen that $|\alpha(t)|$ will exponentially decay to zero in a Markovian environment when $\gamma_0\rightarrow\infty$ as shown by black dotted curve, but decreases more slowly to zero when $\gamma_0$ is smaller and larger memory time.  Specifically, the system dynamics of the non-Markovian case when $\gamma_0=5$ is very similar to that of Markovian case.  

\subsection{Oscillator motion with quantum control}
The decoherence induced by environment can be greatly suppressed by applying a time-dependent LEO control, $H_{LEO}=C(t)a^\dagger a$, where $C(t)$ is the function describing the external pulses operating only on the system of interest. In the following, we will consider three types of pulse control, and analyze the key parameters which determine the performance of quantum control.

We first apply a sequence of normal rectangular pulses added to the system.  These are characterized by $C_1(t)=\omega_1$ for $nT < t \leq nT+\Delta ~~(n\geq0$), and otherwise $C_1(t)=0$. Here, $\omega_1$ is the pulse strength, $T$ represents the time period, and $\Delta$ denotes the duration of the pulse (the width). The detailed solution is provided in Appendix~\ref{rectangular_control}.

Compared to the case without quantum control shown in Fig.~\ref{Fig1}(a), Fig.~\ref{Fig1}(b) indicates that by applying successive rectangular pulses the decoherence of the harmonic oscillator can be remarkably suppressed in a non-Markovian environment. Generally, it is easier to eliminate the decoherence with longer memory time, that is, the smaller $\gamma_0$. Such phenomenon can be interpreted in the following way. The single-mode harmonic oscillator $A$ is coupled to a non-Markovian Ornstein-Uhlenbeck process, which is equivalent to that $A$ is coupled resonantly to the other harmonic oscillator $B$ that is damped in a Markovian environment~\cite{Chaos2005}. The linewidth of the oscillator $B$ is proportional to $\gamma_0$. For small $\gamma_0$, the trap frequency of the oscillator $A$ is effectively renormalized such that two oscillators $A$ and $B$ become off-resonant when applying a series of fast pulses operating on $A$, resulting in the great suppression of decoherence of $A$. While for larger $\gamma_0$, i.e., larger linewidth of oscillator $B$, it's more difficult to make $A$ and $B$ off-resonant. That is, it's more difficult to suppress decoherence with larger $\gamma_0$. As can be seen from Fig.~\ref{Fig1}(b), the decoherence can be completely removed when $\gamma_0=0.1$. When $\gamma_0$ gets larger, although the suppression of decoherence is not perfect, the decoherence can also be slowed down. Interestingly, even for large $\gamma_0=5$ which indicates that the system dynamics is very close to the Markovian case $\gamma_0\rightarrow \infty$ as shown in Fig.~\ref{Fig1}(a), we find that the rectangular pulse control is still effective at suppressing the decoherence as shown in Fig.~\ref{Fig1}(b). Whereas for $\gamma_0\rightarrow \infty$ shown by black dotted curve, the time evolution would not be modulated with any control such that it is identical to the Markovian case given in Fig.~\ref{Fig1}(a). 
\begin{figure}[b]
\centering
\includegraphics[width=0.48\columnwidth]{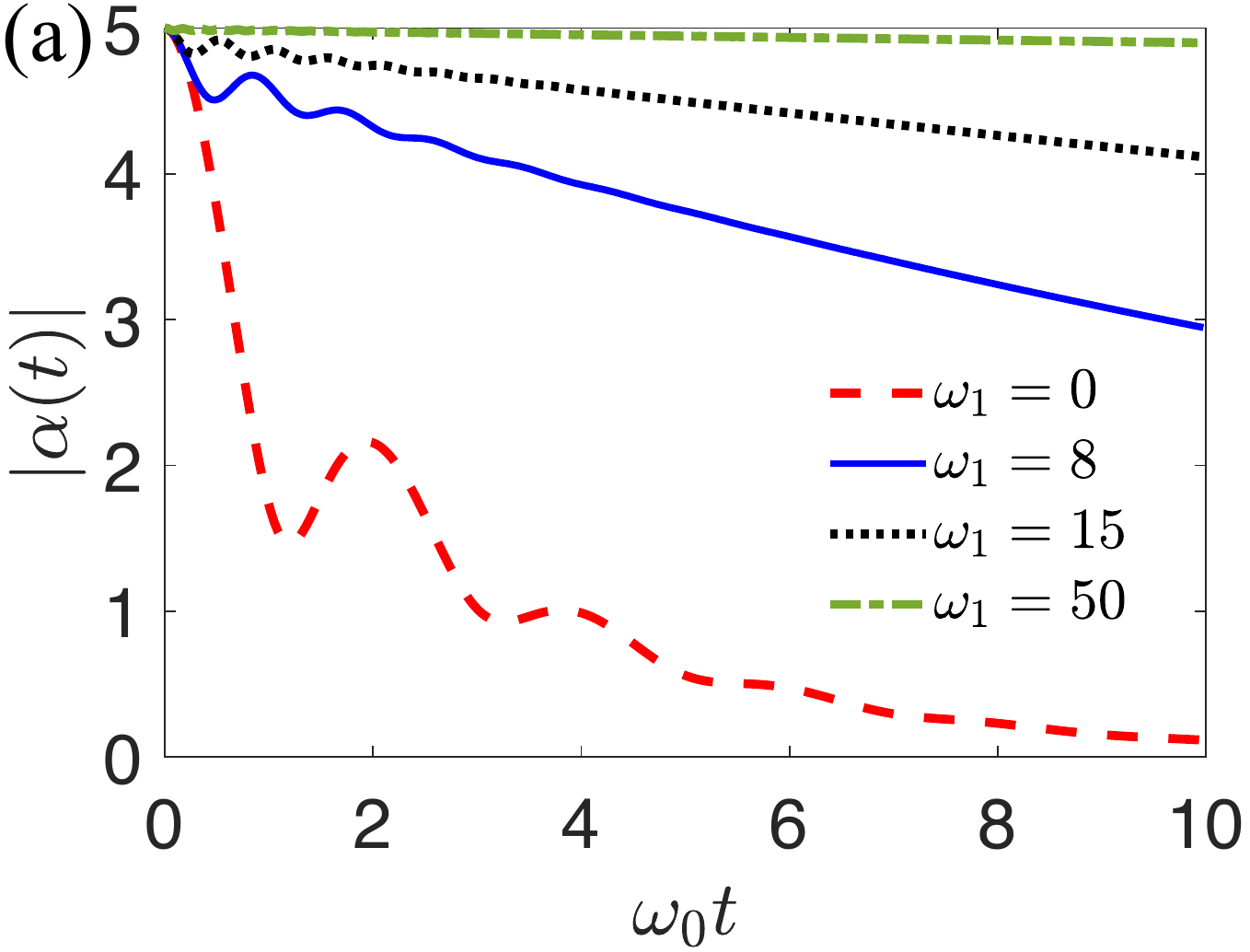}
\includegraphics[width=0.48\columnwidth]{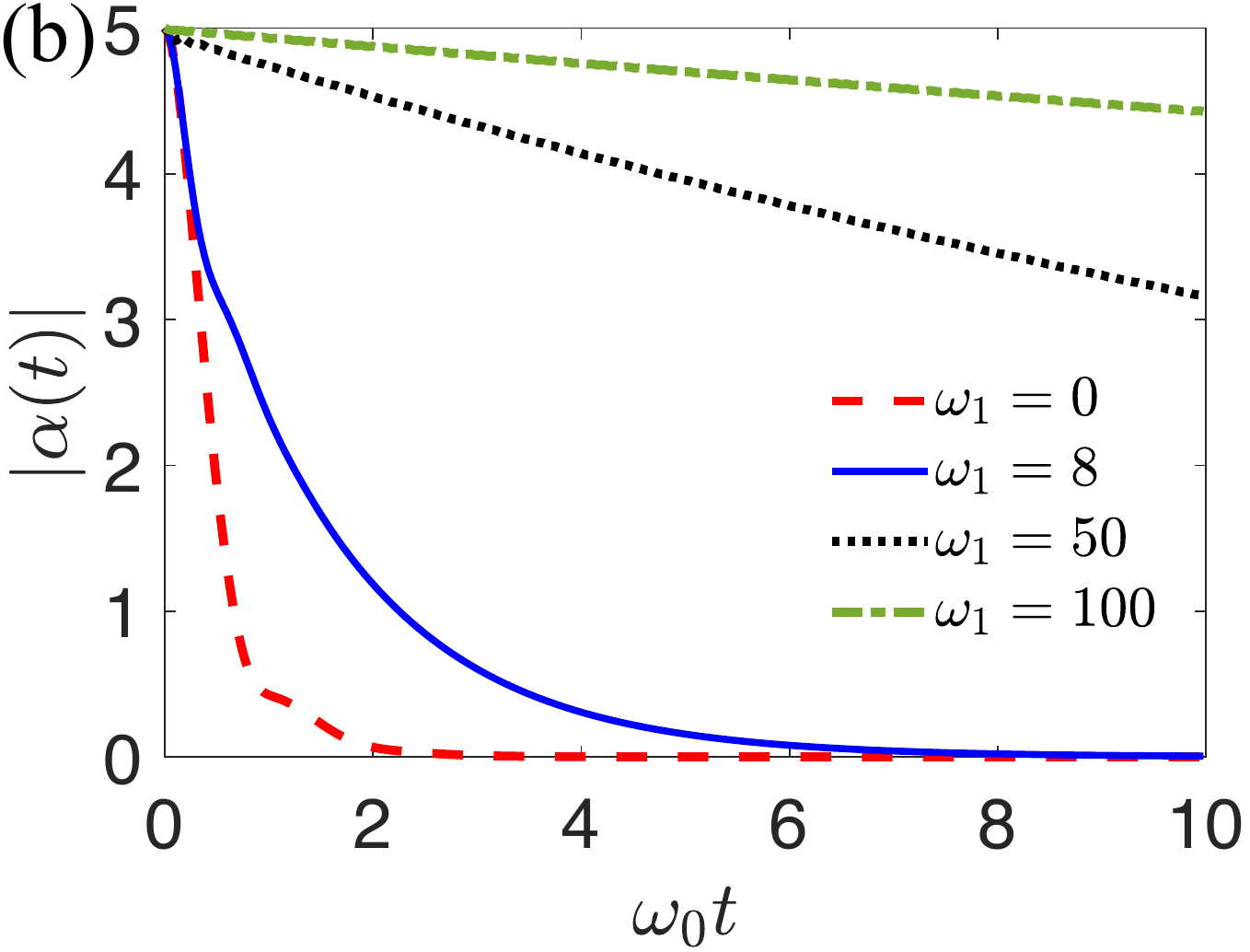}
\includegraphics[width=0.48\columnwidth]{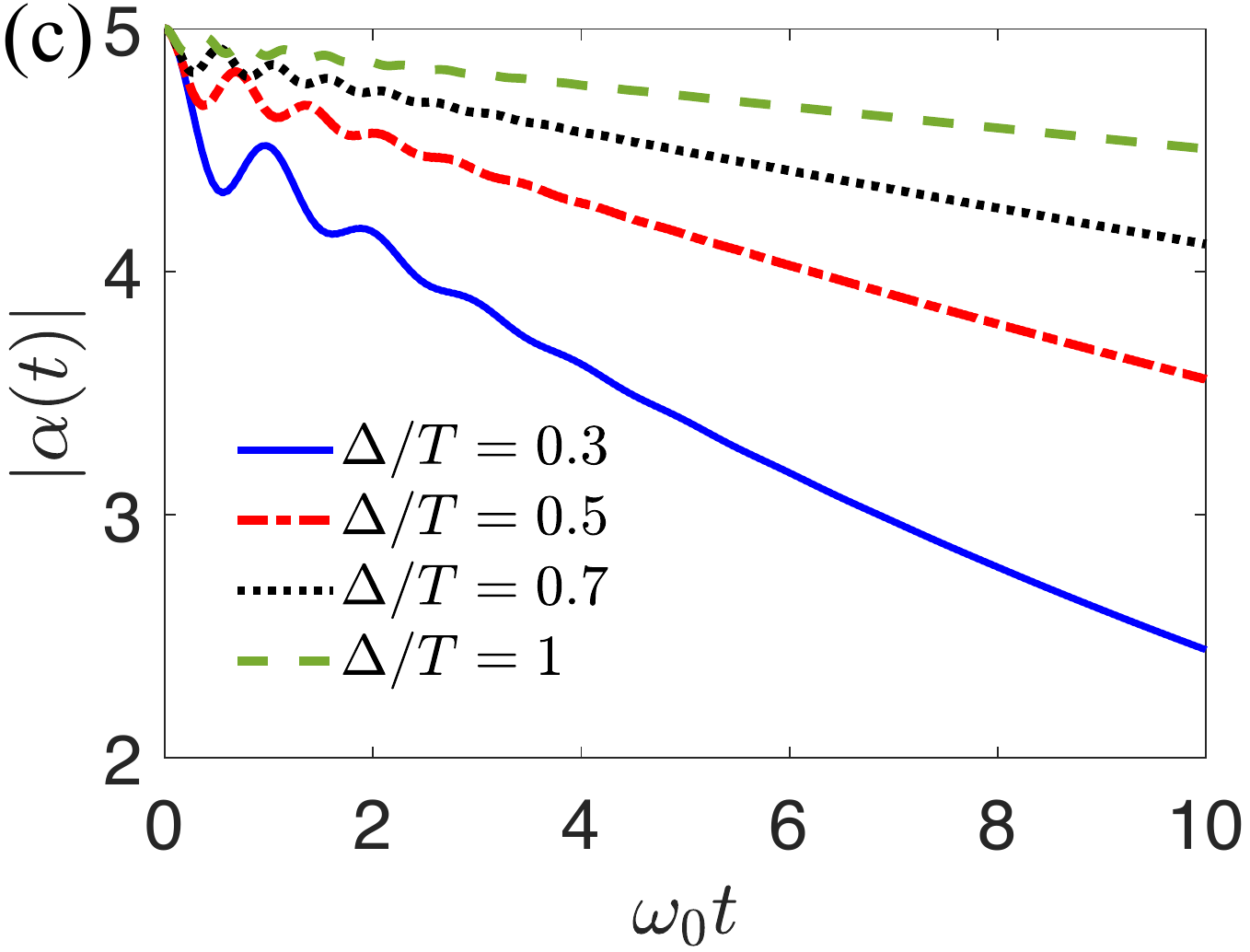}
\includegraphics[width=0.48\columnwidth]{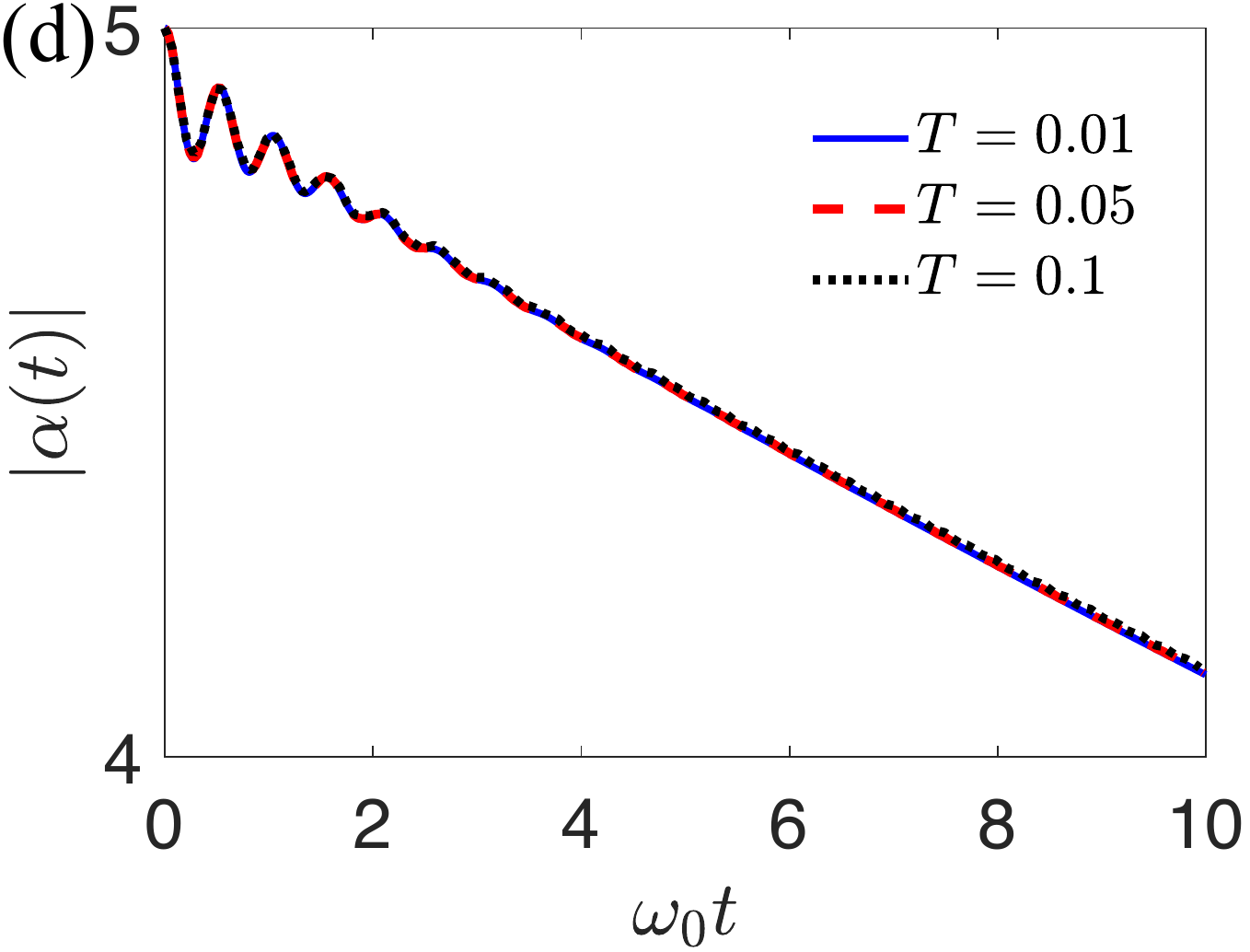}
\caption{The time evolution of $|\alpha(t)|$ for different pulse amplitudes $\omega_1$ when (a) $\gamma_0=1$ and (b) $\gamma_0=5$ with $T=0.05,~\Delta/T=0.7$; for different ratios of pulse duration and period $\Delta/T$ with fixed $T=0.05$ (c), and for different time periods $T$ with fixed ratio $\Delta/T=0.7$ (d) when $\gamma_0=1,~\omega_1=15$. Other parameters are $\alpha(0)=5,~\omega_0=1,~\Gamma=5$.}
\label{Fig2}
\end{figure}

The effectiveness of quantum control can be further improved by increasing the pulse strength $\omega_1$, as depicted in Figs.~\ref{Fig2}(a)-\ref{Fig2}(b). For the case of $\gamma_0=1$ shown in Fig.~\ref{Fig2}(a), $|\alpha(t)|$ nearly remains constant and the desired system is perfectly decoupled from the environment when the pulse amplitude is increased to $\omega_1=50$. By contrast, for the case of $\gamma_0=5$ shown in Fig.~\ref{Fig2}(b), a larger pulse strength is required to overcome the decoherence if the dynamical behavior is more analogous to that of Markovian process. Figures~\ref{Fig2}(c)-\ref{Fig2}(d) show that the ratio between the pulse duration and period $\Delta/T$ is very crucial, while the time period itself is not relevant for the effectiveness of quantum control when $T$ is not very large. It is expected that there is an accelerated decline of coherent amplitude with decreasing $\Delta/T$, as shown in Fig.~\ref{Fig2}(c). The coherent amplitudes with various short time periods $T=0.01,~0.05,~0.1$ for the same ratio $\Delta/T=0.7$ appear to evolve in a similar fashion as plotted in Fig.~\ref{Fig2}(d). This agrees with previous observations that the effectiveness of quantum control is only determined by the integral of the pulse sequence over time if the ratio of duration time and period is suitable~\cite{JingJunPRL2015}.
\begin{figure}
\centering
\includegraphics[width=0.96\columnwidth]{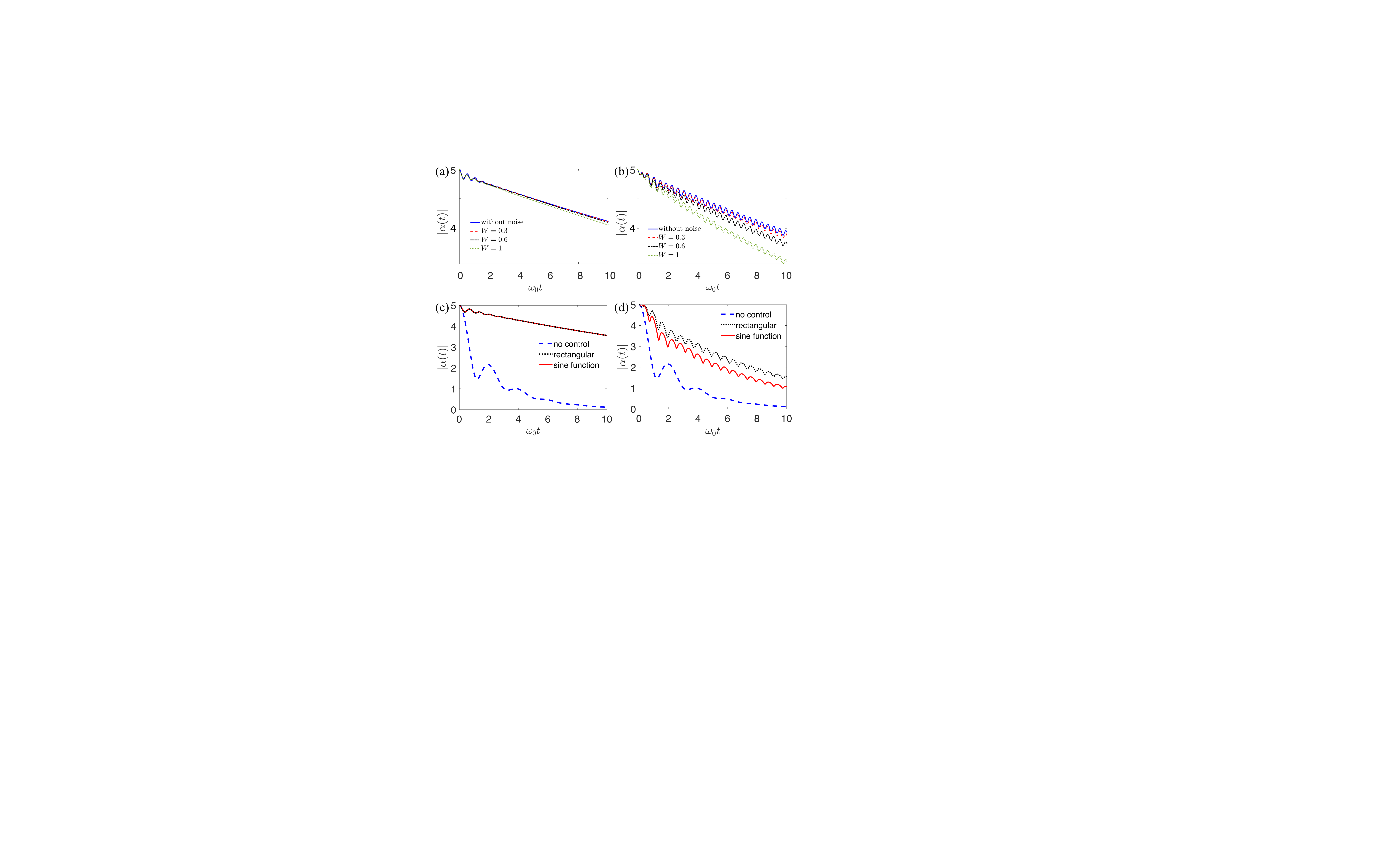}
\caption{(a-b) The time evolution of the expectation value $|\alpha(t)|$ for different amplitudes of Gaussian noise ($\mu=0,~\sigma=1$) with different time periods $T$ in (a) $T=0.05$, (b) $T=0.4$. (c-d) The comparison of time evolution of coherent amplitude $|\alpha(t)|$ for the case without control, under rectangular control and sine function control when (c) $T=0.05$ and (d) $T=0.6$ with $\omega_2=\pi\omega_1/2,~\Delta/T=0.5$. Other parameters are $\alpha(0)=5,~\omega_0=1,~\Gamma=5,~\gamma_0=1,~\omega_1=15, ~\omega_2=\pi\omega_1/2,~\Delta/T=0.5$.}
\label{Fig3}
\end{figure}

In practice, the pulses are noisy and not rectangular due to stochastic quantum fluctuations and environmental noise.  Here we consider a general Gaussian noise~\cite{JingJunPRA2014, JingJunPRL2015, WangPRA2018Jun}, $C_1(t)\rightarrow C_1(t)[1+Wn(t)]$, where $W$ describes the noise strength and $n(t)$ denotes the Gaussian noise with average value $\mu$ and standard deviation $\sigma$. The integral of the pulses remains unchanged within the same time range when taking noise into account, consequently, the effects of quantum control in two cases should be identical in principle. It should be emphasized that the time period of pulses should not be too large, otherwise, the conclusion may differ. We present in Figs.~\ref{Fig3}(a)-\ref{Fig3}(b) the influence of Gaussian noise on the dynamics of the harmonic oscillator with different time periods. It is worth noting that when the pulses are fast enough, $T=0.05$ for example, the noise doesn't significantly change the effects of quantum control as shown in Fig.~\ref{Fig3}(a). Whereas, when the time period is larger, such as $T=0.4$ shown in Fig.~\ref{Fig3}(b), the evolution is more susceptible to the additional noise compared to the case with shorter time period even though the integrals of pulses are the same.

Now consider a sine function pulse control $C_2(t)=\omega_2 \sin{(2\pi t/T)}$ when $nT < t \leq nT+\Delta$~~($n\geq0$), and otherwise $C_2(t)=0$, where the time duration parameter $\Delta$ and time period $T$ are the same as for the regular rectangular pulses, and $\omega_2$ denotes the pulse amplitude. To investigate whether the effectiveness of quantum control depends on the details of the pulses under the condition of possessing the same integral over time, we choose  the pulse strength such that $\omega_2=\pi\omega_1/2$ and $\Delta/T=0.5$. We compare the dynamics of $|\alpha(t)|$ for the case without control, with rectangular control, and with sine function control, respectively, for a small time period $T=0.05$ in Fig.~\ref{Fig3}(c) and a relatively large $T=0.6$ in Fig.~\ref{Fig3}(d). In Fig.~\ref{Fig3}(c), the evolution of rectangular control and sine function control are all the same, while in Fig.~\ref{Fig3}(d), there is a significant difference between these two kinds of controls. We conclude that, when the pulses are relatively fast, the effectiveness of the quantum control depends only on the integrals of pulses over time rather than the details of the successive pulses.  

\begin{figure}
\centering
\includegraphics[width=0.48\columnwidth]{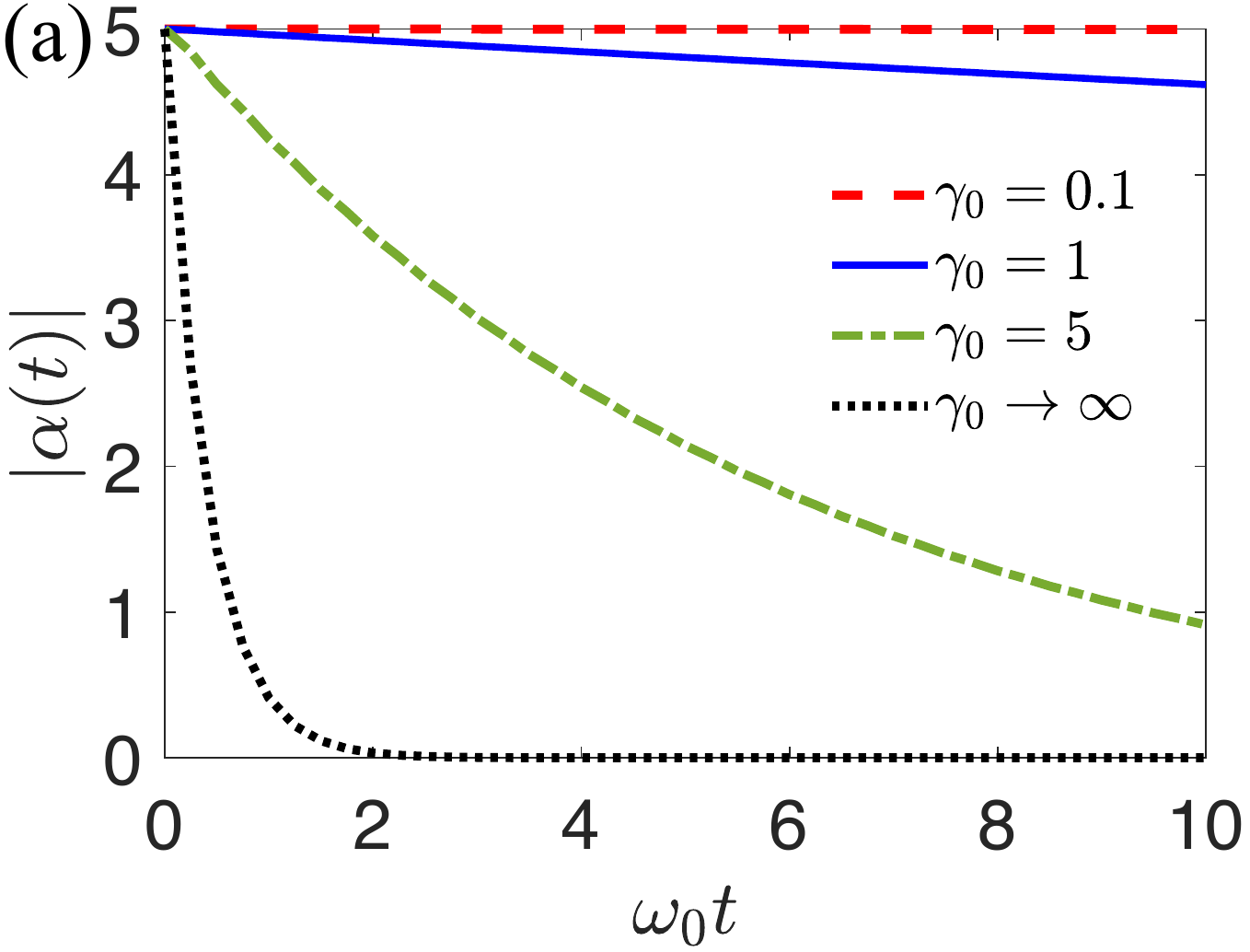}
\includegraphics[width=0.48\columnwidth]{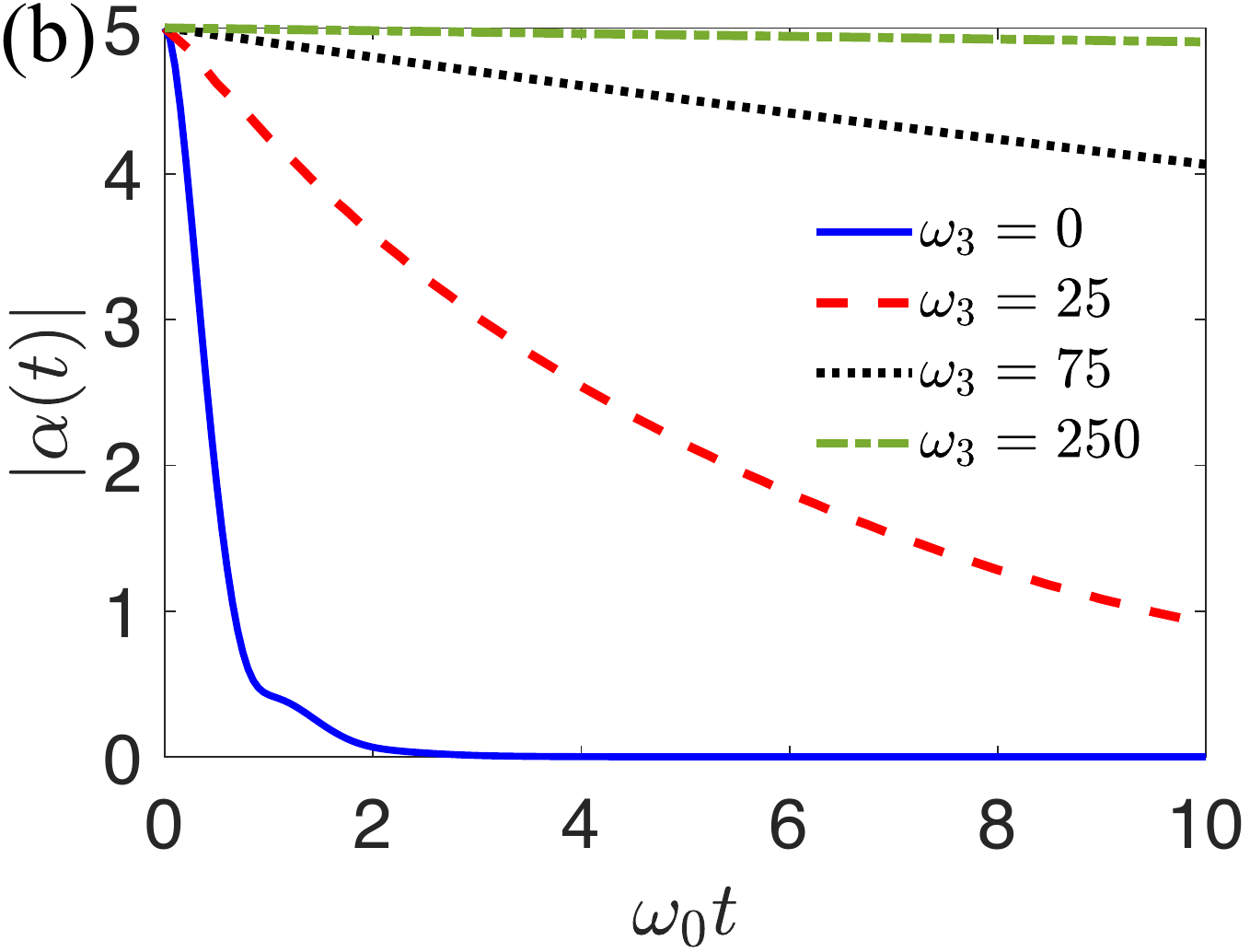}
\caption{(a) The coherent amplitude $|\alpha(t)|$ as a function of time for different $\gamma_0$ with zero-energy control. Other parameters are $\alpha(0)=5,~\omega_0=1, ~\Gamma=5,~\omega_3=25,~T_3=0.5$. (b) For $\gamma_0=5,~T_3=0.5$, the time evolution of $|\alpha(t)|$ for different pulse amplitudes $\omega_3$ when zero-energy pulses are applied. }
\label{Fig4}
\end{figure}

Finally, consider a practical pulse control described by $C_3(t)=\omega_3$ for $nT_3 < t \leq (n+1/2)T_3$ and $C_3(t)=-\omega_3$ for $(n+1/2)T_3 < t \leq (n+1)T_3~~(n\geq0)$, where $\omega_3$ is the pulse strength, and $T_3$ represents the time period. The average value of the energy is zero after a complete control period~\cite{WangPRA2018Jun, WangPRA2018Dec, PyshkinSR2016}. It has been illustrated that such zero-energy control can assist in accelerating holonomic quantum computation~\cite{PyshkinSR2016}. When adopting rectangular pulses or sine function pulses, the average control frequency may be much larger than the bare frequency of the target system. However, the average frequency is zero because of the consecutive sign changes under the control. This  clarifies that the adiabatic speedup effects are not caused by an effective energy increase~\cite{WangPRA2018Dec}. For the same situation, we will show that the decoherence can be greatly inhibited even though affected by zero-energy-cost pulse control, which may eliminate the misunderstanding that the control pulses are only for increasing the effective bare frequency of the oscillator. Compared to the case without any control shown in Fig.~\ref{Fig1}(a), we show in Fig.~\ref{Fig4}(a) that by applying a series of zero-erengy-change pulses, the dynamics of the harmonic oscillator can be remarkably slowed down when the process is non-Markovian. Particularly, when $\gamma_0=0.1$, the coherent amplitude $|\alpha(t)|$ remains constant and the effects of environment can be completely eliminated as shown by the red dashed curve. Moreover, when $\gamma_0$ is large, improving the pulse strength, $\omega_3$, is an effective method to further suppress the decoherence as described in Fig.~{\ref{Fig4}}(b). When the pulse strength is increased to $\omega_3=250$ indicated by green dash-dotted curve, $|\alpha(t)|$ remains nearly unchanged.

\section{Conclusion}\label{Conclusion}
In summary, we have found leakage elimination operators that suppress the decoherence of a single-mode harmonic oscillator coupled to a reservoir of many independent oscillators.  We developed the nonperturbative version of exact LEO theory for the harmonic oscillator by applying the Heisenberg-Langevian method without making approximations.  This is significantly different from previous studies that assumed extremely strong and fast pulses. Moreover, this method is still  applicable for a finite temperature and strong coupling strength. By applying three kinds of consecutive pulses on the desired harmonic oscillator, it was shown that the expectation value of the annihilation operator remains constant over time and the adverse effects induced by the environment can be eliminated. We also found that the effectiveness of LEO control is independent of the fluctuations of the pulse amplitude and the detailed shape of the pulses. It is primarily determined by the integral of the pulse sequences when the time period is relatively small. We expect this work to provide an effective quantum error suppression method for encoded qubits using high-dimensional trapped-ion mechanical oscillator such as in Ref.~\cite{Fluhmann2019}.

We would point out that the proposed dynamical decoupling method is only capable of suppressing the motional decoherence effect for non-Markovian environment. The heating effect described within the framework of Markovian process induces the one-way information loss into the environment which doesn't permit the external pulses to efficiently preserve the coherence. Luckily, for experiments based on trapped ions, the heating of ions can be significantly suppressed either by resolved sideband cooling or by sympathetic cooling with a different species of atomic ions~\cite{NaturePhotonics2017}. With cooling of ions, the quantum motional decoherence due to the coupling with the surrounding environment is one of the remaining dominant factor for the experiment encoding a logical qubit into the state of a trapped ion mechanical oscillator~\cite{Fluhmann2019}. The present method is capable to suppress such decoherence effect in non-Markovian environment. Note that the motional decoherence could be also induced by the dephasing due to trap frequency fluctuations, which is fairly different in nature and may not be directly mitigated by the proposed method, but possible in principle by appropriately choosing other LEO.

\begin{acknowledgments}
This work is supported by the National Key R$\&$D Program of China (Grants No. 2016YFA0301302 and No. 2018YFB1107200), the National Natural Science Foundation of China (Grants No. 11622428, No. 61675007, and No. 11975026), the Key R$\&$D Program of Guangzhou Province (Grant No. 2018B030329001), Beijing Natural Science Foundation (Grant No. Z190005), the Basque Government (Grant No. IT472-10), and the SpanishMICINN (No. FIS2012-36673-C03-03).  MSB was supported by NSF, MPS (Grant No. PHYS-182087).
\end{acknowledgments}

\appendix\label{APPENDIX}

\section{Oscillator motion without quantum control}\label{no_control}
In this Appendix, we present a detailed general solution of the harmonic oscillator with the case of no control.

While defining $\alpha(t)=e^{-(\gamma_0-i\omega_0)t}u(t)$, Eq.~(\ref{alpha_A}) in the main text becomes
\begin{equation}
\dot{u}(t)-(\gamma_0-i\omega_0)u(t)+\frac{\Gamma\gamma_0}{2}\int_0^t u(t')dt'=0,
\end{equation}
with initial condition $u(0)=\alpha(0)$. We can write the general solution as $u(t)=(A_3e^{A_1t}+A_4e^{A_2t})\alpha(0)$ with the coefficients satisfying the following conditions
\begin{eqnarray} \label{coefficient_1}
&& A_1^2-(\gamma_0-i\omega_0)A_1+\Gamma\gamma_0/2=0, \nonumber \\
&& A_2^2-(\gamma_0-i\omega_0)A_2+\Gamma\gamma_0/2=0, \nonumber \\
&&A_2A_3+A_4A_1=0, \nonumber \\
&&A_3+A_4=1.
\end{eqnarray}
We thus obtain the general solution of the harmonic oscillator as
\begin{equation}
\alpha(t)=e^{-(\gamma_0-i\omega_0)t}(A_3e^{A_1t}+A_4e^{A_2t})\alpha(0).
\end{equation}

\section{Oscillator motion with rectangular control}\label{rectangular_control}
In this Appendix, we give an analytical expression of the harmonic oscillator with the case of rectangular control.

For the regular rectangular control pulses, by solving the dynamical equation, we obtain $\alpha(t)=U_{n}(t)\alpha(nT)$ when $nT \leq t \leq nT+\Delta$ ($n \in \mathbb{N}$) with time evolution operator $U_{n}(t)=e^{-(\gamma_0-i c_1)t}(A_{3n}e^{A_{1n}t}+A_{4n}e^{A_{2n}t})$ and $c_1=\omega_0+\omega_1$. Similarly, when $nT+\Delta< t \leq (n+1)T$, $\alpha(t)$ can be easily expressed as $\alpha(t)=U_{n}'(t)\alpha(nT+\Delta)$,
where the time evolution operator $U_{n}'(t)=e^{-(\gamma_0-ic_2)t}(A_{3n}'e^{A_{1n}'t}+A_{4n}'e^{A_{2n}'t})$ with $c_2=\omega_0$. 
For simplicity, we rewrite the evolution operators as $U_{1,n}=U_{n}(nT+\Delta) =e^{-(\gamma_0-ic_1)(nT+\Delta)}[A_{3n}e^{A_{1n}(nT+\Delta)}+A_{4n}e^{A_{2n}(nT+\Delta)}]$ and $U_{2,n}=U_{n}'[(n+1)T] =e^{-(\gamma_0-ic_2)(n+1)T}[A_{3n}'e^{A_{1n}'(n+1)T}+A_{4n}'e^{A_{2n}'(n+1)T}]$, for $t=nT+\Delta$ and $t=(n+1)T$ ($n\in\mathbb{N}$), respectively.
The coefficients $A_{1n},A_{2n},A_{3n}$, and $A_{4n}$ are defined by
\begin{eqnarray} \label{coefficient_21}
&&\frac{A_{3n}}{A_{1n}}e^{A_{1n}nT}+\frac{A_{4n}}{A_{2n}}e^{A_{2n}nT}=-\frac{2e^{(\gamma_0-ic_2)nT}}{\Gamma\gamma_0U_{2,n-1}}\frac{dU_{n-1}'(t)}{dt}\bigg{|}_{t=nT}, \nonumber \\
&&A_{3n}e^{A_{1n}nT}+A_{4n}e^{A_{2n}nT}=e^{(\gamma_0-ic_1)nT},\nonumber \\
&& A_{1n}^2-(\gamma_0-ic_1)A_{1n}+\Gamma\gamma_0/2=0, \nonumber \\
&& A_{2n}^2-(\gamma_0-ic_1)A_{2n}+\Gamma\gamma_0/2=0.
\end{eqnarray}
And the coefficients $A_{1n}',A_{2n}',A_{3n}',A_{4n}'$ satisfy the following conditions
\begin{eqnarray} \label{coefficient_222}
&&\frac{A_{3n}'}{A_{1n}'}e^{A_{1n}'(nT+\Delta)}+\frac{A_{4n}'}{A_{2n}'}e^{A_{2n}'(nT+\Delta)} \nonumber \\
&&~~~~~~~~~~~~~~~~~~~=-\frac{2e^{(\gamma_0-ic_1)(nT+\Delta)}}{\Gamma\gamma_0U_{1,n}}\frac{dU_{n}(t)}{dt}\bigg{|}_{t=nT+\Delta}, \nonumber \\
&&A_{3n}'e^{A_{1n}'(nT+\Delta)}+A_{4n}'e^{A_{2n}'(nT+\Delta)}=e^{(\gamma_0-ic_2)(nT+\Delta)},  \nonumber \\
&& A_{1n}'^2-(\gamma_0-ic_2)A_{1n}'+\Gamma\gamma_0/2=0, \nonumber \\
&& A_{2n}'^2-(\gamma_0-ic_2)A_{2n}'+\Gamma\gamma_0/2=0.
\end{eqnarray}
After numerous iterations, $\alpha(t)$ can be reformulated as
\begin{equation}
\alpha(t)=\left\{
\begin{array}{lcl}
 U_n(t)\alpha(0),~~~~~~~~~~~~~~~~~~~~~~~~{0 < t \leq \Delta,n=0}&&\\
U_n'(t)U_{1,0}\alpha(0), ~~~~~~~~~~~~~~~~~~{\Delta < t \leq T,n=0}&&\\
U_n(t)U_{2,n-1}U_{1,n-1}U_{2,n-2}\cdots U_{2,0}U_{1,0}\alpha(0),   \\
~~~~~~~~~~~~~~~~~~~~~~~~~~~~{nT < t \leq nT+\Delta,n>0}\\
U_n'(t)U_{1,n}U_{2,n-1}U_{1,n-1}U_{2,n-2}\cdots U_{2,0}U_{1,0}\alpha(0),  \\
~~~~~~~~~~~~~~~~~~~~~{nT+\Delta < t \leq (n+1)T,n>0}.
\end{array} \right.
\end{equation}

\bibliography{Control_of_a_harmonic_oscillator}
\end{document}